\documentclass[acmsmall,screen]{acmart}
\usepackage[T1]{fontenc}

\AtBeginDocument{%
  }

\setcopyright{acmcopyright}
\copyrightyear{2024}
\acmYear{2024}
\acmDOI{XXXXXXX.XXXXXXX}
\acmConference[]{}
\acmSubmissionID{}

\acmPrice{15.00}
\acmISBN{978-1-4503-XXXX-X/18/06}
\usepackage{multirow}

\begin{document}

\title{Detecting Continuous Integration Skip : A Reinforcement Learning-based Approach}

\author{Hajer Mhalla}
\email{hajer.mhalla.1@ulaval.ca}
\affiliation{%
  \institution{Laval university}
  \city{Quebec}
  \country{Canada}
}

\author{Mohamed Aymen Saied}
\email{mohamed-aymen.saied@ift.ulaval.ca}
\affiliation{%
  \institution{Laval university}
  \city{Quebec}
  \country{Canada}
}

\begin{abstract}
The software industry is experiencing a surge in the adoption of Continuous Integration (CI) practices, both in commercial and open-source environments. CI practices facilitate the seamless integration of code changes by employing automated building and testing processes. Some frameworks, such as Travis CI and GitHub Actions have significantly contributed to simplifying and enhancing the CI process, rendering it more accessible and efficient for development teams. 
Despite the availability these CI tools , developers continue to encounter difficulties in accurately flagging commits as either suitable for CI execution or as candidates for skipping especially for large projects with many dependencies. Inaccurate flagging of commits can lead to resource-intensive test and build processes, as even minor commits may inadvertently trigger the Continuous Integration process.
The problem of detecting CI-skip commits, can be modeled as binary classification task where we decide to either build a commit or to skip it. This study proposes a novel solution that leverages Deep Reinforcement Learning techniques to construct an optimal Decision Tree classifier that addresses the imbalanced nature of the data.
We evaluate our solution by running a within and a cross project validation benchmark on diverse range of Open-Source projects hosted on GitHub which showcased superior results when compared with existing state-of-the-art methods.\\

\end{abstract}



\maketitle

\section{INTRODUCTION}

Software development practices have advanced and evolved considerably over the years due to the growing demand for more robust products and faster release cycles.
However, as projects become more complex, the risk of human error causing software failures grows alarmingly. 
Continuous integration (CI) is one of the key modern practices used in software development to address this issue \cite{duvall2013ci}. CI proposes that code changes are regularly integrated into a central repository, and automated tests are run to ensure that the code remains in a releasable state.

While this practice offers many benefits such as a faster failure detection, reduced cost, and improved code quality \cite{Vasilescu2015ciquality}, it can incur significant computational resource cost to frequently run and test new versions of the software \cite{luo2017cifactors}. For these reasons, adopting CI can be very expensive and can result in long build times which hinders the development process \cite{ghaleb2019citime}.

Several researchers have attempted to address this issue by proposing solutions such as prioritizing tests and optimizing the build process \citep{Shweta2014buildopt, Haghighatkhah2018testpriority}. Recently, it has been demonstrated  that builds with minor modifications are more likely to pass without errors \cite{abdalkareem2021rule}, therefore omitting them will dramatically lower the cost without sacrificing the benefits of Continuous Integration. This is why, we aim in this research to optimize the continuous integration process by skipping the execution of redundant tests and builds. This problem, called CI-skip, is framed as a classification task wherein the objective is to determine whether a CI process should be executed or skipped.

Nonetheless, explainability stands as a primary constraints when proposing a solution for this problem, developers require insights into  the logic governing build skipping to ensure the integrity of their systems \cite{preece2018explainability}. This is why decision tree models play an important role in the proposed solutions for the CI-skip problem due to its interpretable nature. 
However, decision tree models have limitations, including a tendency to overfit and performs poorly on imbalanced datasets, as it is biased towards the majority class \cite{Truica2017dtimbalance}.

While several solutions, such as resampling techniques \cite{Abdalkareem2021ciskipml} and ensemble methods \cite{Hassan2017randomforest}, have been suggested to address the imbalanced data problem when using decision trees, they have demonstrated shortcomings specifically within the domain of CI-skip.

  Our CI-skip solution utilizes an alternative Decision Tree building algorithm that employs Deep Reinforcement Learning \cite{wen2022rldt} to dynamically optimize the decision tree structure, leading to an improved CI-skip detection.

In order to measure the performance of our approach, we apply both a within-project and cross-project evaluation processes where we compare our approach with different state-of-the-art approaches that tackle the CI-skip problem. In a  within-project evaluation, we train and evaluate the approach on a single project. For the cross-project evaluation, we train on multiple projects and evaluate on a different project.  Furthermore, to gain insights into the decision-making process of our classification model, we provide a feature analysis to identify the attributes that have the most significant influence on the CI-skip decision. Finally, to validate the applicability of our solution,  we tested our solution on Github projects that use Github Actions as a CI/CD framework after adding workflow-level features.

 Compared to other state-of-the-art solutions, our new solution has significantly outperform them and demonstrated an ability to generalize well to unseen projects.

In this paper, we make the following contributions:

\begin{itemize}
    \item A novel approach to handling imbalanced data in the context of CI-skip using Reinforcement Learning in conjunction with Decision Trees.
    \item An empirical evaluation of our approach on real world projects from Travis CI framework and a comparison with state of the art techniques.

    \item An extension of the evaluation of our tool to encompass Github actions workflow features, broadening the scope of applicability and assessing its effectiveness in a diverse set of CI environments.

\end{itemize}

The remainder of this paper is organized as follows. In section \ref{sec2} we review existing literature and research related to the CI-skip problem and we identify gaps and limitations in the current state-of-the-art techniques. In section \ref{sec3}, we describe our approach in detail by explaining how we formulated the problem, the workflow of our approach and the details of its components. Next, we present the design of our experimental study in section \ref{sec4} and the achieved results in section \ref{sec51}. Finally, we discuss the threats to validity of our research in section \ref{sec5}. We conclude by encapsulating our key findings, and delineating avenues for future research.


\section{RELATED WORK }\label{sec2}
\paragraph{}

Over the past decades, software engineering research has addressed numerous challenges across various stages of the software lifecycle. Nonetheless, the rapid evolution in the IT industry, coupled with the exponential growth of new technologies \cite{vayghan2021kubernetes}, including APIs \cite{saied2015could, shatnawi2018identifying, mujahid2021toward}, containers \cite{vayghan2019kubernetes}, microservices \cite{sellami2022improving, almarimi2019web, sellami2022hierarchical, saidani2019towards},  cloud and virtualization, has intensified the demands on software development \cite{benomar2015detection}  and deployment \cite{vayghan2019microservice, vayghan2018deploying} practices to effectively embrace this paradigmatic transition. This prompted continuous scrutiny of established techniques \cite{saied2015could1} and results of software engineering research \cite{saied2020towards, saied2018improving}, fostering exploration into AI and ML-driven approaches to address software engineering challenges across domains such as software reuse \cite{gallais2020api}, recommendation systems \cite{saied2016automated}, mining software repositories \cite{saied2020towards}, software data analytics and pattern mining  \cite{saied2018towards, huppe2017mining, saied2016cooperative, saied2015mining}, program analysis and visualization \cite{saied2015observational, saied2015visualization}, cloud-based testing, Edge-Enabled systems \cite{mouine2022event}, microservices architecture \cite{sellami2022combining},  mobile applications, and DevOps.

\paragraph{}

In this section, we will focus on the use of AI-based techniques to optimize the CI pipeline. Abdalkareem et al. \cite{abdalkareem2021rule} proposed a rule based approach in order to skip commits in the CI build process. These rules were derived from an extensive manual study and involving multiple open-source projects. These rules include, the type of change performed in each commit (i.e., does the commit modify source code, modify the formatting of the source code etc.) and the type of files modified in each commit.
In another work, Xianhao et al. \cite{XianhaoJinRules} observe that these CI-Skip rules might not provide the expected level of safety, their paper proposes PreciseBuildSkip, a collection of CI-Run rules that can complement the rules in \cite{abdalkareem2021rule} to make them safer.

Similarly, Kawrykow et al. \cite{Kawrykow2011rule} proposed a six rule method for detecting the commits that should be skipped which were defined as non-essential changes. Documentation updates, simple type adjustments, and rename-induced modifications are among the rules stated by the authors. Based on their analysis, these non-essential commits reached up to 15\% of the changes. However, this approach does not focus on a CI environment and does not include many of the reasons for which a commit should be skipped such as those specified in \cite{abdalkareem2021rule}.
Moreover, Abdalkareem et al. \cite{Abdalkareem2021ciskipml} attempted to improve upon their prior work through the rule-based approach by designing a method that employs a Decision Tree model. The proposed approach managed to increase the performance in a within-project validation setting but at the cost of a significantly worse performance in the cross-project validation setting.

Leveraging optimization algorithms presents itself as an alternative avenue to address the CI-skip Decision Tree problem.

In this context, Saidani et al. \cite{saidani2022ciskip}, inspired by the success of evolutionary algorithms in dealing with search-based software engineering tasks \cite{Harman2012SBSE}, proposed to formulate the problem of CI skip detection as a search-based problem. Using a tree-based representation of an IF-THEN CI-skip rule, they adapted a genetic algorithm to find optimal detection rules.
Their approach consists in generating a binary tree representation of detection rules and applying a genetic algorithm to find a detection rule that maximizes the true positive rate while minimizing false positives.
However, this approach comes with its own set of challenges, primarily stemming from the vast number of potential feature combinations and their associated threshold values, resulting in a large and complex search space to be explored.\\
On the other hand, build failure detection problem focuses on identifying whether a commit build will result in a failure or success based on its commit features.
The CI-skip detection task and the build failure detection tasks can be seen as quite similar, as they both are essentially binary classification problems that make use of similar commit-level features.

In this context, another approach that made use of a similar genetic approach was introduced by Saidani et al. \cite{saidani2022lstm}  where they focused on predicting failing builds in a CI context. They employed Long Short-Term Memory (LSTM) to capture the temporal correlations in historical CI build data and predict the outcome of the next CI build within a given sequence.
Furthermore, considering that, finding the adequate model configuration (hyperparameters) to use may be a challenging task. This is why they made use of a genetic algorithm which evolves a set of potential LSTM hyperparameters.
However, LSTM models, while effective for capturing temporal dependencies, suffer from scalability limitations, particularly when applied to real world projects. 

Another approach proposed in \cite{Xianhao2}, developed a tool called SmartBuildSkip that works in two phases. Initially it runs a machine learning predictor whether a build is likely to succeed and based on this prediction, it either skips or execute the build. Whenever it observes a failing build, it determines that all subsequent builds will fail. Consequently, it continues with the build process until it eventually observes a successful build, at which point it reverts to prediction mode once again.

Moreover, an effort by Xia and Li \cite{Xia2017buildfail}, for instance, to predict the failure of CI builds considered the problem as a binary classification problem, leveraged 9 Machine Learning classifiers and compared their performance under Area Under Curve (AUC) and F1-score on both a cross-validation and online scenario using commit level features extracted from 126 open-source projects using Travis CI and found that the Random Forest Machine Learning classifier was the model that yields the highest performance.

Additionally, Luo et al. \cite{Luo2017buildfail} conducted a comparative analysis of several machine learning algorithms, including Support Vector Machine (SVM), Decision Tree, Random Forest, and Logistic Regression (LR) on the build failure detection on the TravisTorrent dataset \cite{Beller2017TravisTorrent}. The results suggest that the logistic regression and SVM models achieved the best performance. However, SVM is considered a black box model, making it more challenging to deploy in a developement environment compared to models like Decision Trees and Logistic Regression.

Hassan and Wang \cite{Hassan2017randomforest} similarly made use of the TravisTorrent dataset by defining a feature selection process and using an implementation of the Random Forest algorithm in order to identify build failure without running the process. Remarkably, their approach yielded impressive results, achieving an F1-measure of 87\% when evaluated across various Continuous Integration (CI) environments.

Ni and Li \cite{ni2017ensemble} tackled the CI build failure detection issue by analyzing the most useful prediction features and utilizing the AdaBoost algorithm which outperformed standard decision tree models and a NaiveBayes algorithm.

\section{PROPOSED APPROACH}\label{sec3}
\subsection{Problem formulation}
Ultimately, our goal is to determine whether a commit should trigger the project build and test execution or if it can be safely skipped altogether. This problem can be modelled as a binary classification problem, where we assign a label having two possible values to a commit.

Given a set of features that encapsulate  a commit's metadata, our proposed approach will return whether the commit should be skipped or not. 
Our tool will extend  the commit message by adding the \textit{"[CI SKIP]"}  tag if the decision indicates that the commit can be skipped. 

In this work, a decision tree (DT) based classification technique was considered for two main reasons.
First, previous research \cite{Abdalkareem2021ciskipml} has shown that DT-based approaches deliver promising results in a CI-skip context.
Second, decision tree models are interpretable. Indeed, whereas other Machine Learning models such as Support Vector Machines or the Multi-Layer Perceptron produce probability-based black boxes, for tree-based models' classification, we can identify which features are the most important indicator of whether a commit should be skipped or not. This may be done by analyzing the path followed within the tree structure, recording the nodes traversed during the decision-making process.
However, a major concern with the CI-skip usage and hence any related dataset is data imbalance. As highlighted by the work done by Abdalkareem and al. \cite{abdalkareem2021rule}, CI builds are often executed , resulting in a substantial disparity between the available classes of skipped and executed commits. 

Learning an optimal decision tree is a challenging task as it is known to be NP-complete \cite{DTNP-complete}. The search space for the tree is vast and, therefore, cannot be explored thoroughly. To address this challenge, heuristic methods have been developed for learning decision trees, with greedy search being the most widely used method. While it is simple to understand and implement, greedy search only considers immediate information gain in the current step and makes locally optimal decisions at each node. 
Moreover, for imbalanced datasets, the splitting criteria (e.g. Information Gain) is skew sensitive \cite{Truica2017dtimbalance} which means that, as the prior probability of class is used to calculate a node’s impurity degree, the splitting criteria become biased towards the majority class.

We adapted an alternative Decision Tree building algorithm that leverages Deep Reinforcement Learning to address the issue of imbalanced dataset specifically in the context of handling CI-skip commits. As mentioned in \cite{wen2021rldt} the reward function was designed based on the evaluation metrics commonly used in imbalanced classification tasks. This strategic choice ensures that the agent receives guidance pertinent to CI skip detection throughout the training process.

\subsection{Approach overview and background}
\subsubsection{\textbf{Workflow} }
\begin{figure*}[htbp!]
\centering    
\includegraphics[width=1 \linewidth, height=0.3\textwidth]{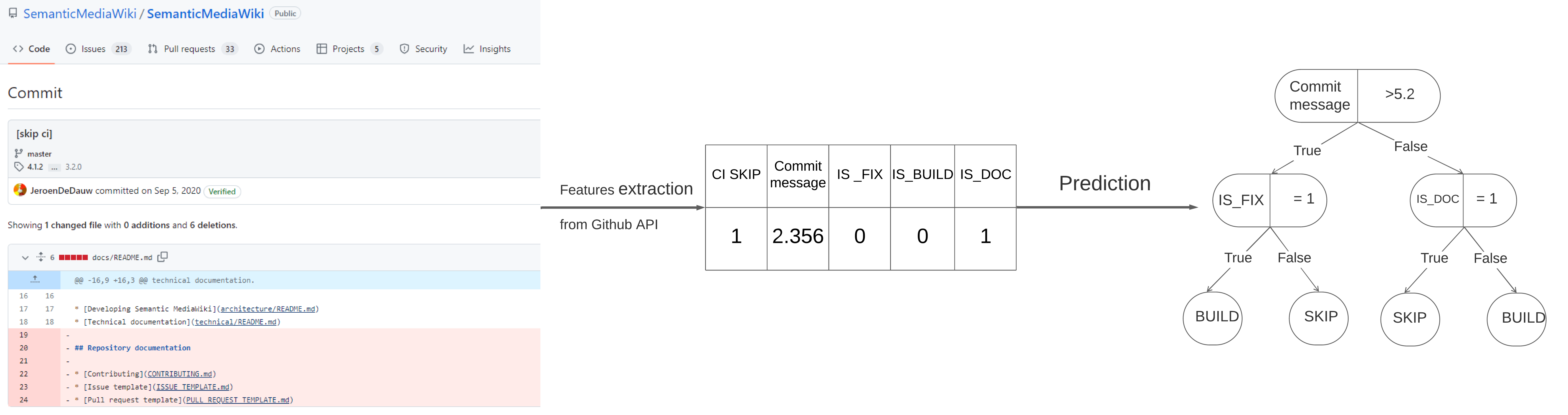}
\caption{The workflow of the proposed approach applied to a commit sampled from our dataset }
\label{fig:workflow}
\end{figure*}
Fig. \ref{fig:workflow} provides an overview of the fundamental workflow of our approach. When a commit is triggered, we extract pertinent commit metadata, which serves as the training data for our model (e.g. Commit\_message, is\_fix, etc.). Using these attributes, we employ a Decision Tree model to classify whether the build should be skipped or not. In case the classification result is skipping the build, we append the \textit{"[CI SKIP]"} flag to the commit message. As such, the CI tool will skip generating the new build and testing it.

The code sample shown in Fig. \ref{fig:workflow} is an example extracted from the Open-Source project SemanticMediaWiki \footnote{https://github.com/SemanticMediaWiki/SemanticMediaWiki} which was included in the TravisTorrent dataset \cite{Beller2017TravisTorrent}. This serves as an example of a commit for a minor change  that would not have any effect on the output of the CI build process. The attributes tables shown in the figure provide a glimpse of the attributes used in our research to train the model. A comprehensive list of these attributes is elaborated upon section \ref{sec4}.
\subsubsection{\textbf{Decision Tree }}

\begin{figure*}[htbp!]
\centering    
\includegraphics[width=0.6 \linewidth, height=0.4\textwidth]{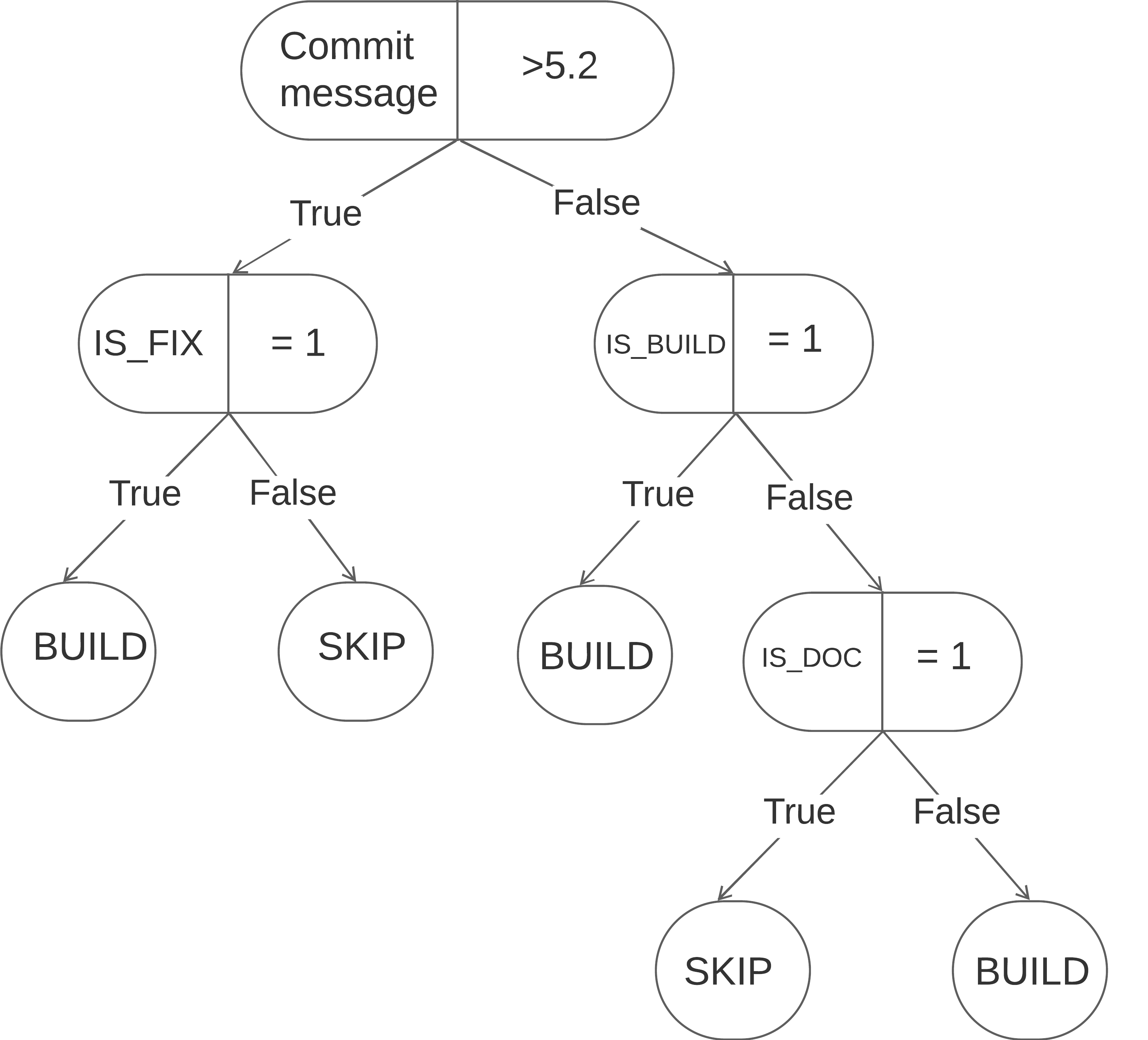}
\caption{An example of a CI-skip decision tree}
\label{fig:decision_tre}
\end{figure*}

 In machine learning, the decision tree can be represented as a binary tree where each node has two main components: the attribute and the threshold. Fig. \ref{fig:decision_tre} showcases an example of this representation. The 4-depth tree has 4 nodes that are each mapped to an attribute and threshold couple. The threshold controls the node's decision based on the value of the commit. As such, based on the code sample shown in Fig.\ref{fig:workflow}, the decision tree will follow the dashed line path resulting in the classification "Skip".
In most problems, the construction of a decision tree can be thought of as a greedy algorithm that selects a locally optimal attribute at each decision node to split the data into sub-nodes. This process is repeated until a leaf node is reached which cannot be further divided.
\subsubsection{\textbf{Deep Reinforcement Learning }}
In general, the core components in a reinforcement learning problem are the agent and the environment. The agent learns to make decisions and optimize its actions based on the feedback received from the environment.

\begin{figure}[htbp!]
\centering    
\includegraphics[width=0.9\linewidth]{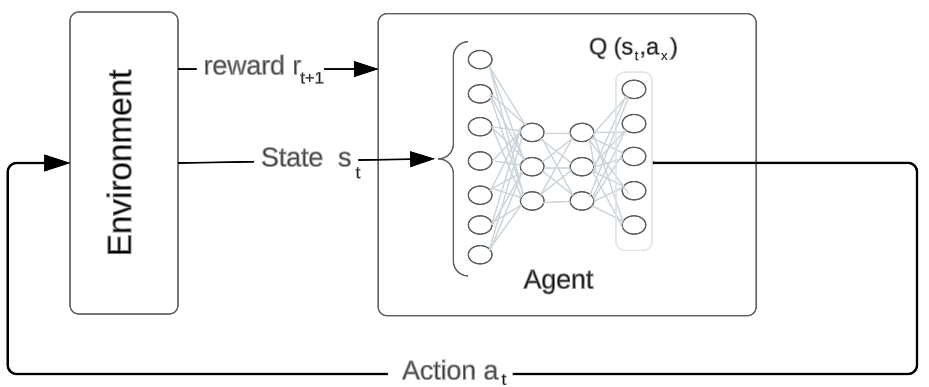}
\caption{Agent-Environment Interaction in Deep Q-Learning}
\label{fig:rl}
\end{figure}

Fig. \ref{fig:rl} provides an abstraction of the concept. The process starts with an initial state $S_0$. At time step $t$, the state $S_t$ is a representation of the environment. Given this state, the agent selects an action $a_t$ based on it policy. This policy dictates the actions that the agent takes as a function of the agent’s state within the environment. \\
Deep Q-Learning \cite{dqnDeepmind} is a deep reinforcement learning technique, where the agent is composed of a Neural Network to estimate the Q-values. Q values are used to determine how good an Action \textit{a}, taken at a particular state \textit{s}, and they are used to sample an action from a policy $p_t$. The action $a_t$, is applied to the environment to generate the next state $S_{t+1}$ and a reward value $R_{t+1}$. The reward serves as feedback to guide the agent's behavior,and the agent's objective is to learn a policy that maximizes the cumulative reward it receives over time. Afterwards, a new cycle starts and the process continues until a defined stopping criteria has been reached (e.g. a predefined maximum number of steps, a terminal state).
An episode refers to a sequence of interactions between an agent and its environment that begins with an initial state and ends when the agent reaches a terminal state in an episodic formulation.\\
As such, in the following section, we formulate the building process of the decision tree as a Markov decision process that can be solved using a deep reinforcement learning algorithm and we present its different components and their interactions. 
\subsection{Building the decision tree}
\subsubsection{\textbf{The state of the environment}}

The decision tree in our case has a fixed structure throughout the training process. It contains $N=2^d-1$ nodes where $d$ is the depth of the tree which is a hyper-parameter that is defined before starting the training process. Each node in the tree contains a single attribute and a threshold value that change during training.

The state $s_t$ needs to be fed to each neural network of the agent. $s_t$ needs to capture information about both the attribute and threshold values of all nodes in the decision tree at timestep \textit{t}.  One possible approach to  represent the state is  an $2*(2^N) + 1$-size vector that encode respectively the attributes and thresholds of nodes with the id of the node to be created obtained by following a breadth-first traversal of the tree structure.

However, when more complex models are needed (i.e. larger trees with a longer depth $d$), the state space grows exponentially. Therefore, we implement tree-based convolution in order to reduce the size of the state representation.
\\Each tree node consisting of an attribute $k$ to a threshold $x_k$ is represented by a one hot vector containing the threshold at the index of the given attribute: $$[0, \dots, 0, x_k, 0, \dots, 0]$$
Hence, at each convolution step, we average the vector representations of each node with its left and right child's therefore, reducing the depth of the tree by one layer.\\
Then, a one-way flatten operation is performed on the convolution's output to pool all features to a 1-dimensional vector that can be fed to the neural networks.\\
Finally, a component is added to the vector to indicate which node is being modified at each step and it is obtained by following a breadth-first traversal of the tree.

\subsubsection{\textbf{The reward}}

After each episode, we apply the new decision tree to classify the training set. The predicted results $\hat{Y}$ and the ground truth $Y$ are used to calculate 2 classification scores, which in our work represent the F1-score and the Area Under Curve (AUC) score (refer to section \ref{sec4} for more details).
Unlike \cite{wen2021rldt} we employ the decision tree to classify only in the terminal node instead of predicting at every step t.\\
Next, the reward $r_m$ is calculated as the difference between two consecutive values of the classification metric $s$ before and after running the episode \textit{m}.
\begin{equation} \label{eq:reward}
r_m = s_m - s_{m-1}
\end{equation}

A positive reward $r_m$ indicates that the action improves the
performance of the classification model, whereas a negative reward
decreases it. This way, during training, the agent is encouraged
to take actions that improve the classification performance.
Furthermore, for this process to account for the imbalanced nature
of data, we choose to evaluate its performance using the F1-score.

\subsubsection{\textbf{The action space}}
At each episode $m$, we modify and evaluate the decision tree by taking a set of parameterized actions $A$. Each action $a_t$ in the set corresponds to the action taken for the node $t$. It is therefore composed of the discrete  attribute $k$ and its corresponding continuous threshold, $x_k$.
$$
A=\bigcup_{0\le t < N}\left\{a_{t}=\left(k, x_{k}\right)\right\}
$$

The composite action is executed to modify the current node  and the process is repeated in a recurrent fashion.  The agent moves to the newly created child nodes and performs the same decision-making process until the stopping criteria is met.

\subsubsection{\textbf{The agent }}
The agent, consisting of two neural networks, observes the state of its environment $s_t$ and takes an action containing a discrete attribute $k$ and a continuous threshold value $x_k$.\\
To do so, the first model, which is the thresholds-network,  computes a threshold vector $X_t$ containing threshold values for all the attributes at time-step t, given the tree state.
\begin{equation} \label{eq:thresholds}
X_t=\mathbb{Q}_x(s_t;\theta_x)
\end{equation}
Where $s_t$ is the state at time $t$ and $\theta_x$ are the weights of the thresholds-network.
Then, given the tree state $s_t$ and the thresholds vector $X_t$, the second model, which is the attributes-network $\mathbb{Q}_q(s_t, X_t;\theta_q)$, calculates the new Q-values based on the iterative calculation of the Bellman optimality equation.
\begin{equation} \label{eq:qvalues}
[q_1, q_2, \dots, q_K]=\mathbb{Q}_q(s_t, X_t;\theta_q)
\end{equation}
Where $\theta_q$ are the weights of the attribute-network, and $q_i$ represents the q value for the attribute \textit{i}.\\
The action space in this reinforcement learning environment has a parametric representation. It is represented as a tuple of an attribute and a threshold. The attribute belongs to a set of discrete and finite values. On the other hand, the threshold can be potentially a continuous variable. As such, theoretically, there is an infinite number of potential actions to choose from and employing standard Q-Learning approaches is not a valid option. For this reason, Wen and Wu \cite{wen2021rldt} proposed in their approach utilizing a parameterized deep Q-Learning approach, and more specifically, the algorithm MP-DQN \cite{Bester2019mpdqn}.  In our approach, we used the P-DQN \cite{Xiong2018pdqn} instead in order to reduce the computational overhead added in the MP-DQN solution. Both approaches rely on two neural networks. The first network attempts at estimating the parameters (i.e. thresholds) values that would maximize the q-values for their corresponding attributes. The second network acts as a standard Deep Q-learning model that selects the best attribute based on the state and the estimated thresholds.  

Weights of the models are updated using the gradient descent algorithm and the loss functions $L_{\mathrm{x}}\left(\theta_{\mathrm{x}}\right)$ and $L_{\mathrm{q}}\left(\theta_{\mathrm{q}}\right)$ for, respectively, the thresholds-networks and attributes-network. They are measured using the following equations:

\begin{equation} \label{eq:lossx}
L_{\mathrm{q}}\left(\theta_{\mathrm{q}}\right)=\mathbb{E}\left(y_{t}-\mathbb{Q}_{\mathrm{q}}\left(s_{t}, \mathrm{xe}_{t k} ; \theta_{\mathrm{q}}\right)\right)^{2}
\end{equation}

\begin{equation} \label{eq:lossq}
L_{x}\left(\theta_{\mathrm{x}}\right)=\mathbb{E}\left(-\sum_{k=1}^{K} \mathbb{Q}_{\mathrm{q}}\left(s_{t}, \mathrm{xe}_{t k} ; \theta_{\mathrm{q}}\right)\right)
\end{equation}
Where $y_{t}$ is the target value and defined as:
\begin{equation} \label{eq:targets}
    y_{t}= \begin{cases}r_{t} & \text {if } s_{t+1} \text {terminal} \\ \gamma \max _{k \in[K]} \mathbb{Q}_{\mathrm{q}}\left(s_{t+1}, \mathrm{xe}_{t+1, k} ; \theta_{\mathrm{q}}\right) & \text {otherwise }\end{cases}
\end{equation}
Where $r_{t}$ is the reward defined in equation \ref{eq:reward} and $s_{t+1}$ is the state at the step $t+1$.

\begin{figure}[htbp!]
\centering    
\includegraphics[width=1.0\linewidth]{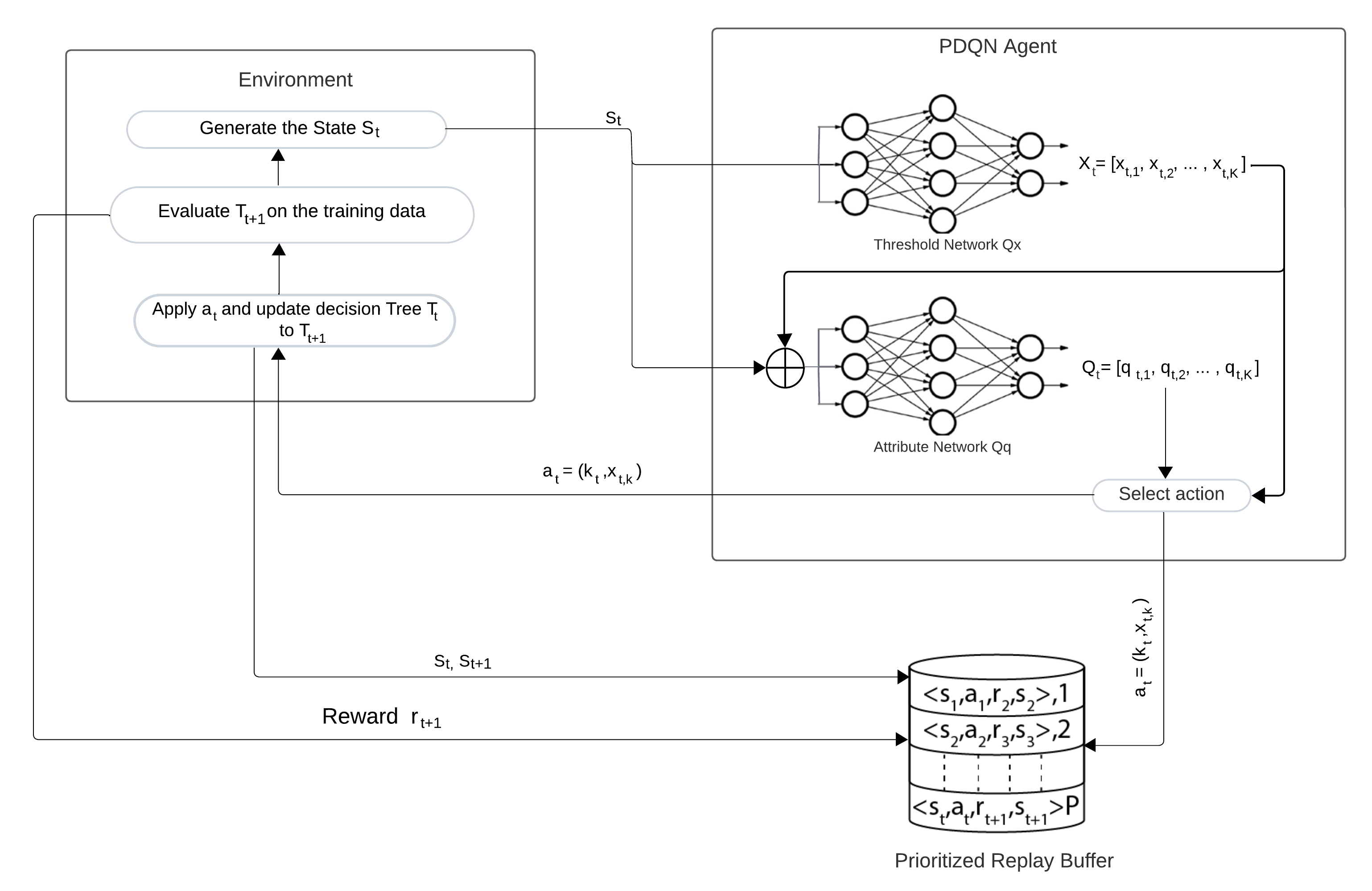}
\caption{The Framework of the proposed solution}
\label{fig:framework}
\end{figure}

\subsubsection{\textbf{Training process}}
  The framework of the proposed approach, as illustrated in Fig. \ref{fig:framework}, outlines the steps required to generate a decision tree.

We first start by building a complete binary tree $T_0$ with a fixed
number of nodes \textit{N} . To do so, we set a random attribute and a
random threshold value for each tree node.
At the same time, the parameters of thresholds-network $\theta_x$ and the attributes-network $\theta_q$ are initialized.\\
Now, at each episode $m$, we reset and rebuild the tree. Each step $t$ corresponds to predicting the feature and threshold of node $t$, $0<t\le N$ and finally a data split.\\
Given the state $S_t$ as input, the agent generates the thresholds vector $X_t$ using the thresholds-network which is concatenated with the state and used by the attributes-network to generate the Q-values.\\
To solve the exploration-exploitation dilemma, we employed an $\epsilon-greedy$ action selection strategy in the feature space. Therefore, at each time step $t$, we randomly sample the action  using the probability $\epsilon$ (exploration) or we select the optimal action  $a_t$ (exploitation) using the following equation:
\begin{equation} \label{eq:actions}
a_{t}= \left(k_{t}, x_{t k}\right) \text { such that } k_{t}=\arg \max _{k \in[K]} \mathbb{Q}_{\mathrm{q}}\left(s_{t}, \mathrm{xe}_{t k} ; \theta_{\mathrm{q}}\right)
\end{equation}
where ${xe}_{t k}$ is a vector where the $k^{th}$ dimension is equal to $x_{tk}$, and everything else is zero.
Afterwards, the environment updates the attribute and threshold value of the $t^{th}$ node of the tree $T_t$ into $k$ and $x_k$ respectively before transitioning from $T_t$ to $T_{t+1}$.\\
Subsequently, we evaluate the decision tree on the test data and the reward $r_t$ is then calculated based on the classification results as defined in equation \ref{eq:reward}.\\
In order to enhance the efficiency and, we incorporate a prioritized experience replay mechanism  \cite{ExperienceReplay}. This technique allows us to mitigate temporal correlations that may exist in the data and improve sample efficiency.

The agent accumulates its experiences, consisting of tuples (state $S_{t}$, action $A_{t}$, reward
$R_{t+1}$, next state $S_{t+1}$, is\_terminal), as it interacts with the environment. These experiences are stored in a data structure known as the replay buffer. Instead of sampling experiences uniformly at random from the buffer during training, the prioritized experience replay assigns priorities to each experience based on their temporal difference (TD) error, which is a measure of how surprising or unexpected the experience was, as a result experiences with low TD errors, which represent situations where the decision tree performed well, are still considered for training, but with lower probabilities.\\

\section{ EXPERIMENTAL STUDY DESIGN}\label{sec4}
\subsection{Research questions}

We designed our experiments to answer four research questions. To examine the effectiveness of the new approach, we performed an empirical study on a benchmark of  21933 commits from 20 projects that use the Travis CI system \cite{Beller2017TravisTorrent}. \\We first consider a within project evaluation in \textbf{RQ1}. Then, we investigate the generalizability of identifying CI skippable commits by applying cross-project evaluation in \textbf{RQ2}, this allows us to assess whether the learnt rules could be successfully applied to different projects beyond the ones used in the training data.
To validate the performance, we compare our results with 2 state-of-the-art
techniques from literature:
\begin{itemize}
    \item  \textbf{SPEA-2} \cite{saidani2022ciskip}: Decision rules obtained using a Multi-objective Evolutionary Search which was shown to be the approach that yields the highest classification score for our task. 
    \item \textbf{DT}: Decision Tree based on Gini Index \cite{Abdalkareem2021ciskipml} to help assess the effectiveness of the RL approach against the greedy approach.

\end{itemize}
One of our main objectives is to deliver interpretable results, that is why in \textbf{RQ3} we analyse the learnt rules and we investigate what features are most important to the decision-making process by  performing a top node analysis.
 Lastly, in \textbf{RQ4} we extended the evaluation of our tool to include commits collected from GitHub Actions, thereby assessing its effectiveness across a diverse set of Continuous Integration (CI) frameworks.

\subsection{Dataset}
To evaluate this novel approach, we initially considered the same dataset used in \cite{saidani2022ciskip}
which comprises data from 15 open-source Java and Ruby projects  that were integrated with the Travis CI system. 
However, during the evaluation process, we identified and removed 3 projects from the dataset due to specific selection criteria mentioned  in \cite{saidani2022ciskip} not being met. To ensure the reliability of our model, the \textit{training} dataset  was constructed to contain a minimum of 200 commits and include at least 10\% of skipped commits on the master branch. For this purpose, we excluded the projects "\textit{future}" and "\textit{parallelc}" due to the small size of their training data. Additionally, we removed the project "\textit{steve}" as only 4.03\% of its commits were identified as skipped.
We extracted 8 additional java and ruby projects using Google Big Query to filter the projects and Github API to extract the features, and we ended up with 21933 commits from 20 projects. The statistics about the new studied projects including the number of commits and the percentage of the skipped commits can be found in Tab.\ref{tab:tabrq31}. 
\begin{table}[htbp]
\caption{Statistics about the new projects}~\label{tab:tabrq31}
\begin{center}
\begin{tabular}{|l|l|l|}
\hline
\textbf{Project} &  \textbf{\#Commits} & \textbf{\% skipped com.} \\

\hline

xylophone&  319 &39\% \\
\hline

jekyll-sitemap & 337 & 20\%  \\
\hline
minimapper & 322 & 16\%  \\
\hline
wisper & 306 & 14\%  \\
\hline
figaro & 378 & 10\%  \\
\hline

mockito  &4215 & 15\%   \\
\hline
rm-notify-gateway & 167 & 16\%   \\
\hline
pl.wrzasq.parent & 815 & 17\%   \\
\hline
\end{tabular}
\label{tab:tabrq31}
\end{center}
\end{table}

\subsection{Commit Features}
\begin{table}[t]
\centering
\caption{\small Features used to train our CI-Skip detection tool extracted from literature \cite{saidani2022ciskip} \cite{Abdalkareem2021ciskipml}
}
  \begin{tabular}{l l }
\hline   Feature & Description  \\
\hline 
\textit{NS} & Number of  modified sub-systems. \\ 
 \textit{ND}  & Number of changed directories. \\
\textit{NF}& Number of changed files. \\ 
\textit{ENTROPY} &Distribution of modified code across each file. \\ 
 \textit{LA}   & Number of added lines.   \\ 
\textit{LD}  & Number of lines of code deleted.    \\
\textit{Day\_week}& Day of the week the commit is performed. \\ 
\textit{CM} & Measures the importance of terms appearing in the commit  message using TF-IDF.\\ 
 \textit{TFC} &Number of Changed files’ types identified by their extension.\\
\textit{CLASSIF}&Feature Addition (1), Corrective (2),Perfective(4), Preventative (5), \\&Non-Functional (6), None (7). \\ 

\textit{IS FIX} &It is a bug fixing commit. \\
\textit{IS DOC} &If commit affects only documentation files (e.g. README, docx). \\
\textit{IS BUILD} &If the commit contains only build files (e.g. make). \\ 
\textit{IS META} &If the commit contains only metadata files.  \\
\textit{IS MERGE} &If it is a merge commit (has more than 2 parents). \\ 
\textit{IS MEDIA} &The commit contains only media files (e.g. mp3). \\  
\textit{IS SRC} &The commit affects only source code files. \\ 
\textit{FRM} &The commit is a formatting of the source code. \\ 
\textit{COM} &The commit modifies only source code comments. \\ 
\textit{PRS} &Number of skipped commits in the 5 past commits. \\ 
\textit{CRS} &Number of Commits that were skipped by the current committer in the 5 \\&past commits. \\ 
\textit{PCR} &If the previous commit was skipped. \\ 
\textit{SC} &If it is the same committer of the last commit. \\ 
\textit{NUC} &Number of Unique Changes. \\
\textit{AGE} &The average time interval between the current and the last time affected files\\& were modified. \\ 
\textit{NDEV} & The number of developers that previously changed the touched file(s). \\  
\textit{LT} &Size of changed files before the commit. \\  
\textit{EXP} &Number of commits made by the developer. \\ 
\textit{SEXP} & Subsystem experience measures the number of commits affecting the modified \\&subsystems. \\ 
\textit{REXP} &Recent experience is the total experience of the developer in terms of commits, \\&weighted by their age\\

\hline

\hline
\end{tabular}
 
  \label{tab:features}
\end{table}

For each commit, the original dataset includes 30 attributes described in Tab.\ref{tab:features}.  After conducting a feature correlation analysis, we identified some features with a high Pearson correlation index ($\geq 80\%$). Although decision tree classifiers naturally handle multicollinearity, our reinforcement learning approach heavily relies on features to define the action space. Highly correlated features might introduce redundancies in the action space, leading to inefficiencies in training and exploration. Consequently, we opted to eliminate the features "ND," "NF," "REXP," and "CRS" and trained our model using the remaining 26 features.

\subsection{Metrics}

In order to review and compare research projects found in the literature, it is essential first to define how we are able to quantify the performance of a solution, and how we should conduct its evaluation. \\
Our problem can be modeled by a binary classification problem, therefore, we can apply the usual metrics used in classification tasks. 
However, we will therefore focus on metrics that are most relevant to imbalanced classification tasks like CI-skipping commits.\\
In all our experiments, we employ the well-known evaluation metrics where we compute the F1-score defined as follows:
\begin{equation}
F1-score=2* \frac{Precision*Recall} {Precision+recall} \in [0,1]
\end{equation}
Where : 
\begin{equation}
Recall= \frac {TP} {(TP+FN)} \in [0,1]
\end{equation}
\begin{equation}
Precision= \frac {TP} {(TP+FP)} \in [0,1]
\end{equation}
We also consider the Area Under the ROC Curve (AUC) values which measures how well the model performs
on the minority and majority classes and it is defined as
follows.
\begin{equation}
AUC = \frac{1+  \frac{TP}{TP +FN} - \frac{FP}{FP +TN}} {2} \in [0,1]
\end{equation}

\section{Results}\label{sec51}

\textit{\textbf{RQ1}. How does our approach perform compared to SOTA techniques in a within project evaluation?}
\\

\textbf{Motivation}
To properly assess the efficacy of our method, it is important to compare it first against state-of-the-art techniques discussed previously in a within-project validation scenario.\\
Therefore, we are considering the case of an existing project with a rich history whose developers want to optimize by adding the automated CI-skip detection feature.

\textbf{Approach}
To answer RQ1, we use the 26 commit-level features to train and evaluate each project apart against SPEA-2and the Decision tree Gini-based classifier.\\
For each project, our model is trained and evaluated on two different chunks of its commit history. We split each dataset into training and test set in proportion of 80\% and 20\% with stratification to ensure that the proportions of labels in each class remained consistent with the original datasets. The
training set is used to build the decision tree and formulate the reward function, while the test set is employed to assess the performance of the final model. \\
However, during the investigation of the SPEA-2 implementation \footnote{https://github.com/stilab-ets/SkipCI}, we had to adopt the same training and test set-based strategy. To ensure a fair comparison we took the initiative to decompile their tool and implement a stratified train test split and recalculate the evaluation metrics. And as we have collected the most up-to-date data for the active projects examined in our work, we proceeded to retrain the decision tree classifier.\\ 
Unlike the DT classifier, we did not perform any oversampling technique  in order to highlight the contribution of the RL policy on handling imbalanced data.\\
To compare the predictive performance of  our solution with other state-of-the-art techniques, we use F1-score and AUC that were previously defined.\\

\textbf{Results}
Table \ref{tab:rq1_results} showcases the results for all projects using the AUC and F1-score metrics in a the within-project evaluation scenario. 
We can observe that our approach managed to achieve an average AUC score of 88\% and median value equal to 88\%, achieved the best results for 17 out of 20 projects, showing a significant average improvement of 9\%. This demonstrates that our approach can achieve the best balance between both minority and majority classes, surpassing all other studied methods.\\
For the F1-score, our solution achieved the best results for 18 out of 20 projects with an average score of 81\% and a median score equal to 84\%.
 These results underscore the effectiveness of the RL-based decision tree in identifying  CI skip commits more accurately.

 However, in the case of the "\textit{SemanticMediaWiki}" project, we encountered a considerable drop in our model's performance. To investigate the issues, we initially trained a Gini-based decision tree. For this specific project, which contains around 8000 commits, the optimal decision tree required a minimum depth of 15. However, when we applied this depth to our RL-based approach, we encountered a scaling issue.
 Due to the nature of our approach, as the state size grows exponentially with the tree depth ($2*(2^N) + 1$), the size of the replay memory also increases significantly, and also a larger neural network is needed to handle this state space. This led to practical challenges in terms of computational resources and memory efficiency, making it difficult to maintain the model's performance with a smallest tree depth.

\begin{table}[H]
\centering
 \caption{\small Performance of RL-based DT vs. other techniques under within-project validation}~\label{tab:rq1_results}
\scalebox{0.75}{
\begin{tabular}{|l|ccc|ccc|}
\hline
& \multicolumn{3}{c|}{AUC}& \multicolumn{3}{c|}{F1-Score} \\ \cline{2-7} 
\multirow{-2}{*}{\textit{\textbf{Project}}} & \multicolumn{1}{c|}{Ours}  & \multicolumn{1}{c|}{SPEA-2} &\multicolumn{1}{c|}{DT}  & \multicolumn{1}{c|}{Ours}& \multicolumn{1}{c|}{SPEA-2}& \multicolumn{1}{c|}{DT} \\ \hline

\textbf{candybar-library}&  
\multicolumn{1}{c|}{\textbf{79}} & \multicolumn{1}{c|}{79}   & \multicolumn{1}{c|}{82}   &

\multicolumn{1}{c|}{\textbf{89}}  &  \multicolumn{1}{c|}{80}  &\multicolumn{1}{c|}{83}
\\ \hline

\textbf{contextlogger}&  
\multicolumn{1}{c|}{\textbf{100}} & \multicolumn{1}{c|}{92}   & \multicolumn{1}{c|}{94}  &

\multicolumn{1}{c|}{\textbf{100}}  &  \multicolumn{1}{c|}{95}  &\multicolumn{1}{c|}{96}
\\ \hline
\textbf{figaro}&  
\multicolumn{1}{c|}{\textbf{83}} & \multicolumn{1}{c|}{72}   & \multicolumn{1}{c|}{65}   &
\multicolumn{1}{c|}{\textbf{60}}  &  \multicolumn{1}{c|}{50}  &\multicolumn{1}{c|}{39}
\\ \hline
\textbf{GI}&  
\multicolumn{1}{c|}{82} & \multicolumn{1}{c|}{87}   & \multicolumn{1}{c|}{\textbf{89}}& 
\multicolumn{1}{c|}{\textbf{87}}  &  \multicolumn{1}{c|}{79}  &\multicolumn{1}{c|}{70}
\\ \hline

\textbf{grammarviz2\_src}&  
\multicolumn{1}{c|}{\textbf{99}} & \multicolumn{1}{c|}{85}   & \multicolumn{1}{c|}{81}  &
\multicolumn{1}{c|}{\textbf{96}}  &  \multicolumn{1}{c|}{67}  &\multicolumn{1}{c|}{77}
\\ \hline
\textbf{groupdate}&  
\multicolumn{1}{c|}{\textbf{82}} & \multicolumn{1}{c|}{66}   & \multicolumn{1}{c|}{73}  &

\multicolumn{1}{c|}{\textbf{64}}  &  \multicolumn{1}{c|}{61}  &\multicolumn{1}{c|}{59}
\\ \hline
\textbf{jekyll-sitemap}&  
\multicolumn{1}{c|}{90}& \multicolumn{1}{c|}{\textbf{98}}   & \multicolumn{1}{c|}{92} &
\multicolumn{1}{c|}{97}  &  \multicolumn{1}{c|}{\textbf{99}}  &\multicolumn{1}{c|}{88}
\\ \hline
\textbf{minimapper}&  
\multicolumn{1}{c|}{\textbf{97}} & \multicolumn{1}{c|}{87}   & \multicolumn{1}{c|}{92}   &
\multicolumn{1}{c|}{\textbf{88}}  &  \multicolumn{1}{c|}{73}  &\multicolumn{1}{c|}{88}
\\ \hline
\textbf{mockito}&  
\multicolumn{1}{c|}{\textbf{92}} & \multicolumn{1}{c|}{82}   & \multicolumn{1}{c|}{91}   &
\multicolumn{1}{c|}{\textbf{79}}  &  \multicolumn{1}{c|}{77}  &\multicolumn{1}{c|}{75}
\\ \hline
\textbf{mtsar}&  
\multicolumn{1}{c|}{\textbf{73}} & \multicolumn{1}{c|}{70}   & \multicolumn{1}{c|}{72}& 
\multicolumn{1}{c|}{\textbf{64}}  &  \multicolumn{1}{c|}{61}  &\multicolumn{1}{c|}{64}
\\ \hline
\textbf{pghero}&  
\multicolumn{1}{c|}{\textbf{88}} & \multicolumn{1}{c|}{76}   & \multicolumn{1}{c|}{87}&
\multicolumn{1}{c|}{\textbf{81}}  &  \multicolumn{1}{c|}{67}  &\multicolumn{1}{c|}{63}
\\ \hline
\textbf{pl.wrzasq.parent}&  
\multicolumn{1}{c|}{\textbf{98}} & \multicolumn{1}{c|}{98}  & \multicolumn{1}{c|}{92}  &
\multicolumn{1}{c|}{\textbf{95}}  &  \multicolumn{1}{c|}{92}  &\multicolumn{1}{c|}{89}
\\ \hline
\textbf{ransack}&  
\multicolumn{1}{c|}{\textbf{85}} & \multicolumn{1}{c|}{72}   & \multicolumn{1}{c|}{83} &
\multicolumn{1}{c|}{\textbf{79}}  &  \multicolumn{1}{c|}{62}  &\multicolumn{1}{c|}{77}
\\ \hline
\textbf{rm-notify-gateway}&  
\multicolumn{1}{c|}{\textbf{89}} & \multicolumn{1}{c|}{64}   & \multicolumn{1}{c|}{78} &
\multicolumn{1}{c|}{\textbf{93}}  &  \multicolumn{1}{c|}{40}  &\multicolumn{1}{c|}{66}
\\ \hline
\textbf{SAX}&  
\multicolumn{1}{c|}{\textbf{98}} & \multicolumn{1}{c|}{88}   & \multicolumn{1}{c|}{87}& 
\multicolumn{1}{c|}{\textbf{94}}  &  \multicolumn{1}{c|}{85}  &\multicolumn{1}{c|}{82}
\\ \hline

\textbf{searchkick}&  
\multicolumn{1}{c|}{\textbf{87}} & \multicolumn{1}{c|}{83}   & \multicolumn{1}{c|}{85}&  
\multicolumn{1}{c|}{\textbf{77}}  &  \multicolumn{1}{c|}{77}  &\multicolumn{1}{c|}{72}
\\ \hline
\textbf{SemanticMediaWiki}&  
\multicolumn{1}{c|}{58} & \multicolumn{1}{c|}{58}   & \multicolumn{1}{c|}{\textbf{62}} &
\multicolumn{1}{c|}{32}  &  \multicolumn{1}{c|}{\textbf{41}}  &\multicolumn{1}{c|}{38}
\\ \hline
\textbf{solr-iso639-filter}&  
\multicolumn{1}{c|}{\textbf{81}} & \multicolumn{1}{c|}{65}   & \multicolumn{1}{c|}{79}&
\multicolumn{1}{c|}{\textbf{77}}  &  \multicolumn{1}{c|}{65}  &\multicolumn{1}{c|}{76}
\\ \hline

\textbf{wisper}&  
\multicolumn{1}{c|}{\textbf{85}} & \multicolumn{1}{c|}{64}   & \multicolumn{1}{c|}{75}&
\multicolumn{1}{c|}{\textbf{67}}  &  \multicolumn{1}{c|}{40}  &\multicolumn{1}{c|}{52}
\\ \hline

\textbf{xylophone}&  
\multicolumn{1}{c|}{\textbf{99}} & \multicolumn{1}{c|}{98}   & \multicolumn{1}{c|}{98}&
\multicolumn{1}{c|}{\textbf{99}}  &  \multicolumn{1}{c|}{99}  &\multicolumn{1}{c|}{97}
\\ \hline

\hline

\textit{Median}      &   \multicolumn{1}{c|}{\textbf{88}} & \multicolumn{1}{c|}{80} & \multicolumn{1}{c|}{85} & \multicolumn{1}{c|}{\textbf{84}}& \multicolumn{1}{c|}{70}& \multicolumn{1}{c|}{75}  \\ \hline

\textit{Average}      & \multicolumn{1}{c|}{ \textbf{88}} & \multicolumn{1}{c|}{79} & \multicolumn{1}{c|}{83} & \multicolumn{1}{c|}{\textbf{81}} & \multicolumn{1}{c|}{71}& \multicolumn{1}{c|}{72} \\ \hline
\end{tabular}
}

\end{table}

\textit{\textbf{RQ2}. How does our approach perform compared to SOTA techniques in a cross-project evaluation ?}\\

\textbf{Motivation} In the  previous research question, we considered the case of  an existing project  with a rich CI-skip history.
Another crucial scenario arises when a development team initiates a new project. Here, historical labeled data is often lacking, posing a challenge for building an effective AI model. In this context, the CI-skip detection tool must be trained using data from other projects that ideally share similar tools and technologies. Cross-project evaluation stands as the state-of-the-art technique to address this data scarcity issue by importing data from other projects.\\

\textbf{Approach}
Similar to previous works\cite{Abdalkareem2021ciskipml, saidani2022ciskip}, we train the classifier based on data from 19 projects and use the reserved project to test the trained DT. This process is repeated for all studied projects. \\
To gain better insights into the performance of our approach, we compare it with the ML techniques used in RQ1 also based on F1-score and AUC.\\
Similar to the previous research question, we adapted SPEA-2 implementation to include a stratified train-test split and we recalculated the evaluation metrics for the test projects. \\
As a result, the training data was constructed using all available data except the test project, ensuring that the model is evaluated on unseen data.\\

\textbf{Results}
Table \ref{tab:rq2_results} showcases the results for all projects and studied approaches using the AUC and F1-score metrics for the cross-project evaluation. 
\begin{table}[H]
\centering
 \caption{\small Performance of RL-based DT vs. other techniques under cross-project validation}~\label{tab:rq2_results}
 \scalebox{0.75}{
\begin{tabular}{|l|ccc|ccc|}
\hline
& \multicolumn{3}{c|}{AUC}& \multicolumn{3}{c|}{F1-Score} \\ \cline{2-7} 
\multirow{-2}{*}{\textit{\textbf{Project}}} & \multicolumn{1}{c|}{Ours}  & \multicolumn{1}{c|}{SPEA-2} &\multicolumn{1}{c|}{DT} & \multicolumn{1}{c|}{Ours}& \multicolumn{1}{c|}{SPEA-2}& \multicolumn{1}{c|}{DT} \\ \hline

\textbf{candybar-library}&  
\multicolumn{1}{c|}{\textbf{70}} & \multicolumn{1}{c|}{69}   & \multicolumn{1}{c|}{66}& 

\multicolumn{1}{c|}{\textbf{82}}  &  \multicolumn{1}{c|}{80}  &\multicolumn{1}{c|}{73}
\\ \hline

\textbf{contextlogger}&  
\multicolumn{1}{c|}{\textbf{99}} & \multicolumn{1}{c|}{89}   & \multicolumn{1}{c|}{89}   &

\multicolumn{1}{c|}{\textbf{99}}  &  \multicolumn{1}{c|}{85}  &\multicolumn{1}{c|}{87}
\\ \hline
\textbf{figaro}&  
\multicolumn{1}{c|}{\textbf{73}} & \multicolumn{1}{c|}{65}   & \multicolumn{1}{c|}{57}  &
\multicolumn{1}{c|}{\textbf{49}}  &  \multicolumn{1}{c|}{44} &\multicolumn{1}{c|}{24}
\\ \hline
\textbf{GI}&  
\multicolumn{1}{c|}{\textbf{91}} & \multicolumn{1}{c|}{72}   & \multicolumn{1}{c|}{68}  &
\multicolumn{1}{c|}{\textbf{82}}  &  \multicolumn{1}{c|}{37}  &\multicolumn{1}{c|}{51}
\\ \hline

\textbf{grammarviz2\_src}&  
\multicolumn{1}{c|}{\textbf{89}} & \multicolumn{1}{c|}{74}   & \multicolumn{1}{c|}{68}   &
\multicolumn{1}{c|}{\textbf{77}}  &  \multicolumn{1}{c|}{49}  &\multicolumn{1}{c|}{53}
\\ \hline
\textbf{groupdate}&  
\multicolumn{1}{c|}{\textbf{79}} & \multicolumn{1}{c|}{66}   & \multicolumn{1}{c|}{73}   &

\multicolumn{1}{c|}{58}  &  \multicolumn{1}{c|}{47}  &\multicolumn{1}{c|}{\textbf{59}}
\\ \hline
\textbf{jekyll-sitemap}&  
\multicolumn{1}{c|}{\textbf{96}} & \multicolumn{1}{c|}{50}   & \multicolumn{1}{c|}{49}  &
\multicolumn{1}{c|}{\textbf{89}}  &  \multicolumn{1}{c|}{0}  &\multicolumn{1}{c|}{21}
\\ \hline
\textbf{minimapper}&  
\multicolumn{1}{c|}{\textbf{94}} & \multicolumn{1}{c|}{73}   & \multicolumn{1}{c|}{56}   &
\multicolumn{1}{c|}{\textbf{87}}  &  \multicolumn{1}{c|}{61}  &\multicolumn{1}{c|}{27}
\\ \hline
\textbf{mockito}&  
\multicolumn{1}{c|}{\textbf{80}} & \multicolumn{1}{c|}{77}   & \multicolumn{1}{c|}{56}   &
\multicolumn{1}{c|}{\textbf{66}}  &  \multicolumn{1}{c|}{56}  &\multicolumn{1}{c|}{27}
\\ \hline
\textbf{mtsar}&  
\multicolumn{1}{c|}{\textbf{61}} & \multicolumn{1}{c|}{60}   & \multicolumn{1}{c|}{59}   &
\multicolumn{1}{c|}{\textbf{52}}  &  \multicolumn{1}{c|}{51}  &\multicolumn{1}{c|}{37}
\\ \hline
\textbf{pghero}&  
\multicolumn{1}{c|}{\textbf{75}} & \multicolumn{1}{c|}{57}   & \multicolumn{1}{c|}{59}   &
\multicolumn{1}{c|}{\textbf{58}}  &  \multicolumn{1}{c|}{46}  &\multicolumn{1}{c|}{35}
\\ \hline
\textbf{pl.wrzasq.parent}&  
\multicolumn{1}{c|}{\textbf{95}} & \multicolumn{1}{c|}{76}   & \multicolumn{1}{c|}{61}   &
\multicolumn{1}{c|}{\textbf{84}}  &  \multicolumn{1}{c|}{58}  &\multicolumn{1}{c|}{34}
\\ \hline

\textbf{ransack}&  
\multicolumn{1}{c|}{\textbf{82}} & \multicolumn{1}{c|}{74}   & \multicolumn{1}{c|}{66}   &
\multicolumn{1}{c|}{\textbf{68}}  &  \multicolumn{1}{c|}{52}  &\multicolumn{1}{c|}{47}
\\ \hline

\textbf{rm-notify-gateway}&  
\multicolumn{1}{c|}{\textbf{82}} & \multicolumn{1}{c|}{55}   & \multicolumn{1}{c|}{59}   &
\multicolumn{1}{c|}{\textbf{76}}  &  \multicolumn{1}{c|}{19}  &\multicolumn{1}{c|}{31}
\\ \hline
\textbf{SAX}&  
\multicolumn{1}{c|}{\textbf{91}} & \multicolumn{1}{c|}{74}   & \multicolumn{1}{c|}{76}   &
\multicolumn{1}{c|}{\textbf{84}}  &  \multicolumn{1}{c|}{49}  &\multicolumn{1}{c|}{65}
\\ \hline
\textbf{searchkick}&  
\multicolumn{1}{c|}{\textbf{83}} & \multicolumn{1}{c|}{71}   & \multicolumn{1}{c|}{67}   &
\multicolumn{1}{c|}{\textbf{70}}  &  \multicolumn{1}{c|}{65}  &\multicolumn{1}{c|}{50}
\\ \hline
\textbf{SemanticMediaWiki}&  
\multicolumn{1}{c|}{\textbf{57}} & \multicolumn{1}{c|}{70}   & \multicolumn{1}{c|}{53}   &
\multicolumn{1}{c|}{\textbf{30}}  &  \multicolumn{1}{c|}{44} &\multicolumn{1}{c|}{25}
\\ 
\hline

\textbf{solr-iso639-filter}&  
\multicolumn{1}{c|}{\textbf{73}} & \multicolumn{1}{c|}{68}   & \multicolumn{1}{c|}{65}   &
\multicolumn{1}{c|}{64}  &  \multicolumn{1}{c|}{\textbf{68}}  &\multicolumn{1}{c|}{64}
\\ 
\hline


\textbf{wisper}&  
\multicolumn{1}{c|}{\textbf{72}} & \multicolumn{1}{c|}{57}   & \multicolumn{1}{c|}{61}   &
\multicolumn{1}{c|}{\textbf{52}}  &  \multicolumn{1}{c|}{26}  &\multicolumn{1}{c|}{32}
\\ \hline

\textbf{xylophone}&  
\multicolumn{1}{c|}{\textbf{86}} & \multicolumn{1}{c|}{77}   & \multicolumn{1}{c|}{70}  &
\multicolumn{1}{c|}{\textbf{84}}  &  \multicolumn{1}{c|}{74}  &\multicolumn{1}{c|}{64}
\\ \hline


\hline

\textit{Median}       & \multicolumn{1}{c|}{\textbf{82}} & \multicolumn{1}{c|}{70} &   \multicolumn{1}{c|}{63} & \multicolumn{1}{c|}{\textbf{73}} & \multicolumn{1}{c|}{67} & \multicolumn{1}{c|}{50} \\ \hline

\textit{Average}      & \multicolumn{1}{c|}{ \textbf{81}} & \multicolumn{1}{c|}{69} &  \multicolumn{1}{c|}{64} & \multicolumn{1}{c|}{\textbf{71}} & \multicolumn{1}{c|}{65} & \multicolumn{1}{c|}{51}  \\ \hline
\end{tabular}}

\end{table}
We can observe that our approach yielded promising results with a median of 82\% and 73\% in terms of AUC and F1-score respectively, and an average of 81\% and 71\% in terms of AUC and F1-score respectively, outperforming SPEA-2 and he Gini-based DT with an average improvement of 15\% in terms of AUC and 13\% in terms of F1-score.\\
Due to the increased size and diversity of the training dataset in the cross-project validation setting, the reinforcement agent will be forced to explore a much larger space in order to learn and find the decision tree parameters that can achieve better results in multiple projects which, as seen by the average AUC values for all approaches, differ in difficulty. For this reason, the reinforcement learning agent will find the decision tree parameters that can generalize across multiple projects and is more robust against distributional shift. On the other hand, SPEA-2 and the Gini-based DT can perform better with smaller datasets but struggle to generalize in larger datasets.
Our agent was able to develop a generalized policy, where the features learnt are flexible and reusable learnt features across all projects unlike the SPEA-2 and the Gini-based DT classifiers.

\textit{\textbf{RQ3}.  What features are most important to detect skippable commits ?}

\textbf{Motivation}
So far, we have evaluated our approach and compared it against other CI skip techniques, but  it is also important to explore and analyse the decision trees in order to understand what factors influenced the decision of skipping a commit. Through this research question, we aim to give better insights to developers and assist them in making the skipping decision and hence improving the model explainability. In the other hand, the feature importance can help researchers better analyse the policy learnt by the RL agent.\\
\textbf{Approach} In order to interpret the classifications performed by our model, we conduct a feature importance analysis where we calculate the importance of a node as the decrease in node impurity weighted by the probability of reaching the node. Notably, the most important features can vary from one project to another, and there is no specific feature that consistently appears in the top node analysis of all studied projects.\\
We first calculate the node probability  which is simply the number of samples that reach the node Nj, divided by the total number of samples and we denote that as W($N_j$). \\
$W(N_j)=w_j \in (0,1]$\\
We calculate the feature importance of $N_j$ using the formula:
\begin{equation}
U(N_j)=w_ji_j^2-W(N_j^l)I^2(N_j^l)-W(N_j^r)I^2(N_j^r)
\end{equation}

Where : 

$I^2 $: The mean squared error,

$N_j^l, N_j^r$: left and right child of node j.

We calculate the features importance for each project in the dataset using the methodology outlined below, then for each project we calculate 
the relevance of a feature as the sum of the nodes' importance where the considered feature occurs. We finally sum the statistics of all trees constructed under within-project evaluation and extract the top-4 factors that influence the CI-Skip decision.\\

\textbf{Results}
\begin{table}[htbp]
\caption{Top features analysis}~\label{tab:rq4_results}
\begin{center}
\begin{tabular}{|c|c|c|}
\hline
\textbf{TOP-4} & \textbf{Project} & \textbf{Importance} \\

\hline 

\multirow{3}{*}{CM} & SAX              & 0.87                \\ \cline{2-3} 
                              & candybar-library         & 0.77                \\ \cline{2-3}  & figaro           & 0.69                \\ \hline
                              
ProjRecentSkip  & ransack  & 0.58    
 \\ \hline     

\multirow{2}{*}{exp} & SearchKick       & 0.55                \\ \cline{2-3} 
                              & pghero           & 0.35                \\ \hline

TFC    & Groupedate & 0.43  \\                  

\hline
\end{tabular}
\label{tab1}
\end{center}
\end{table}

Table  \ref{tab:rq4_results} presents the top-4 features across all projects examined, based on the within-project validation.
It is worth mentioning that the most important features can differ from one project to another, and there is no specific feature that appears in the top node analysis of all studied projects.

Results show that the commit message is the most important feature to detect CI Skip commits, since developers mention the type of modification, hence it is a useful information to decide whether or not to skip a build. This feature proves to be particularly influential in the projects \textit{SAX}, \textit{candybar-library}, and \textit{figaro}, aligning with the findings of  \cite{Abdalkareem2021ciskipml}\cite{saidani2022ciskip}.

\textit{ProjRecentSkip}, which is a feature that indicates the number of recently skipped commits from the last 5 commits, also appears in the top node analysis in the project \textit{ransack}. This result suggests that there is a strong link between
current and previous commits results, so if the recent commits are skipped, it is probable that the next commits will be skipped as well.
Similarly, the developer experience, represented by the number of total  commits made by the developer, is the top feature in the projects \textit{SearchKick} and \textit{pghero}, since experienced developers are more familiar with the CI-Skip concept.

Finally, the type of files changed is also an important indicator since it can reflect the type of change done, this observation aligns with suggestions in \cite{abdalkareem2021rule} and was proved in \cite{Abdalkareem2021ciskipml}.

This ability to identify and utilize diverse features for different projects highlights the adaptability and effectiveness of the RL-based approach in the context of CI-Skip detection.
These results are exciting and proves how the RL agent can manage to learn from scratch, thought trial and error, the same rules extracted manually or learned by SPEA-2 algorithm and Gini-based decision tree, using a simple and unique reward function.

\textit{\textbf{RQ4}. How does our approach perform when trained on Github Actions workflow features?}\\

\textbf{Motivation} Among the various CI/CD tools, Github Actions \cite{GithubActions} has emerged as a prominent choice due to its seamless integration with the popular code-hosting platform, GitHub.
In this research question, we aim to investigate two essential aspects of our approach for CI skip detection. The first objective is to assess how effectively our tool performs when applied to a new framework. While our previous research questions have demonstrated promising results on TravisCI based  projects, it is crucial to validate the tool's performance in a new CI/CD such as Github Actions in order to gain insights into its adaptability and generalization capability.
Secondly, we seek to conduct a comprehensive comparison between two training configurations for our tool. The first  involves training the model solely on commit-level features \textit{CLF}, similar to RQ1. The second configuration  involves incorporating both commit-level and workflow-level features \textbf{CLF+WLF} during training. By conducting this comparison, we aim to determine the impact of leveraging Github Actions workflow features on the tool's performance. This investigation can shed light on whether the additional context provided by workflow features improves our CI skip detection tool performances.\\

\textbf{Approach} 
In this study, we conducted an investigation building upon the approach used in RQ1. Following the same filtering criteria, we selected projects that uses Github Actions and had at least 200 commits, with 10\% of these commits being skipped.

Subsequently, we extracted features from five nominated projects detailed in Tab.\ref{tab:RQ5.0}. We kept the same commit level features and for the workflow-specific attributes, we focused on three key attributes described in Tab.\ref{tab:RQ5.1}.
\begin{table}[htbp]
\caption{Statistics about the Github Actions projects}~\label{tab:RQ5.0}
\begin{center}
\begin{tabular}{|l|l|l|}
\hline
\textbf{Project} &  \textbf{Number of Commits} & \textbf{Percentage of skipped commits} \\

\hline

liquibase-percona&   1109  &18.85\%\\
\hline

conda-forge.github.io&   4163  & 20\%  \\
\hline
wmixvideo-nfe & 2641 & 21.54\%    \\
\hline
django-maintenance-mode & 530 & 14.91\%   \\
\hline
heroku-maven-plugin & 826 &  27.24\%  \\
\hline
\end{tabular}
\end{center}
\end{table}

\begin{table}[htbp]
\centering
\caption{\small Workflow-level Features used to train our CI-Skip detection tool} 
  \begin{tabular}{|l |l| }
\hline   \textbf{Feature} & \textbf{Description}  \\
\hline 
\textit{PBS} &  Result of the last build . \\ 
\hline 
\textit{Fail\_rate}  & The fail rate of the builds by the current 
committer in the past. \\
\hline 
\textit{avg\_exp }  & The average number of builds the committers made
in the project before the \\&current build. \\

\hline

\end{tabular}
  \label{tab:RQ5.1}
\end{table}

The initial step involves training our model using commit-level features with within-project evaluation. Subsequently, we conduct a comparative analysis by training the model on hybrid features, which comprise a combination of both commit-level features and workflow-level features. \\

\textbf{Results} Tab. \ref{tab:RQ5.2} show the results for both, the model trained on commit-level features (CLF) and the model trained on hybrid features (CLF+WLF) for each project. The F1 score and AUC (Area Under the Curve) are the metrics we used for evaluation.

\begin{table}[H]
\caption{\small Results of training on CLF and training on CLF+WLF} 
\scalebox{0.85}{
\begin{tabular}{|l|ll|ll|}

\hline
\multirow{2}{*}{\textbf{Project}}  & \multicolumn{2}{c|}{\textbf{CLF}}      & \multicolumn{2}{c|}{\textbf{CLF+WLF}}  \\ \cline{2-5} 
                           & \multicolumn{1}{l|}{\textit{F1}} & \textit{AUC} & \multicolumn{1}{l|}{\textit{F1}} & \textit{AUC} \\ \hline
\textbf{liquibase-percona}        & \multicolumn{1}{l|}{61}   &   79  & \multicolumn{1}{l|}{72}   & 89    \\ \hline
\textbf{conda-forge.github.io}    & \multicolumn{1}{l|}{78}   &  90   & \multicolumn{1}{l|}{80}   &  93   \\ \hline
\textbf{wmixvideo-nfe}               & \multicolumn{1}{l|}{91}   &   94  & \multicolumn{1}{l|}{95}   &  97   \\ \hline
\textbf{django-maintenance-mode}    & \multicolumn{1}{l|}{59}   &   73  & \multicolumn{1}{l|}{55}   &    79 \\ \hline

\textbf{heroku-maven-plugin}&  \multicolumn{1}{l|}{70}   &  84   & \multicolumn{1}{l|}{72}   & 89    \\ \hline \hline
\textit{Median}&  \multicolumn{1}{l|}{70}   &  84   & \multicolumn{1}{l|}{72}   & 90    \\ \hline
\textit{Average}&  \multicolumn{1}{l|}{72}   &  84   & \multicolumn{1}{l|}{75}   & 88    \\ \hline
\end{tabular}}
\label{tab:RQ5.2}
\end{table}

Upon analyzing the table, it is evident that adding workflow-level features to the model consistently improves its performance across all projects. It is worth noting that the magnitude of improvement varies from project to project, with some projects exhibiting more significant enhancements than others especially for the \textit{liquibase-percona} which is considered a relatively large project.
The average F1-score shows a substantial increase from 61\% to 72\%, while the AUC score improves from 79\% to 89\%. These results demonstrate the overall positive impact of integrating additional Github Actions workflow-level information along with commit-level features during the model training process.

\section{Threats to validity}\label{sec5}
Threats to validity can be internal or external. For internal threats to validity, the observed results and their analysis are based on certain hyper-parameters of the different approaches that were analyzed, including ours. To mitigate this threat, we applied a hyper-parameter tuning process for each project and selected the results that achieved the highest scores.  The second threat to validity is the correctness of the implementation of our approach. In order to mitigate this issue, we extensively tested our implementations and verified the soundness of the achieved results. A critical point to note is that our results reflect the best performance of the agent and not necessarily the convergence value. This distinction is important as it indicates that while the agent performs well under certain conditions, its ability to converge to an optimal solution might differ. One further step towards improvement would be to study the stability issues of the agent to ensure consistent performance across episodes. As for external threats to validity, our approach was evaluated on only 20 Open-Source projects from TravisCI and 5 from Github Actions. In order to achieve a better generalization of our approach, it would be ideal to train and evaluate on a larger number of projects and workflows. To mitigate this issue, we utilized projects with varying scales and workflows.

\section{conclusion}\label{sec6}
In this research, we tackled the problem of detecting the commits within a Continuous Integration pipeline that should be skipped. 
We propose a Decision Tree based model due to its inseparability. We utilize a Reinforcement Learning process to build the Decision Tree in order to account for the imbalanced data within this task.

We evaluate the effectiveness of our approach by conducting both within-project and cross-project
validation and comparing it to best existing methods. In both cases, our approach showcased a greater capacity at generalizing and robustness against distributional shift in cross project evaluationby achieving higher F1-scores and AUC values than all state of the art approaches. 
Regarding the features analysis, we
found that the commit message, the number of recent skip  commits, the committer experience and the number of Changed files’ type sare the most
influential features in CI skip detection.
The DQN-based decision tree represents an interesting case study on the potential of the reinforcement learning algorithms to overcome the problem of imbalanced datasets in software engineering without the need to rebalance the data and sacrifice the DT interpretability.
Finally, o assess the adaptability of our tool across various CI/CD platforms, we collected data from 5 projects utilizing the Github Actions framework. Our experiments revealed that our approach not only performs well on this new framework but also exhibits enhanced performance when incorporating workflow-related features. 
As a future enhancement of our technique, we envision to experiment with the combination of Random forest classifier and DQN where the classifiers are constructed by a set of cooperative agents jointly as proposed in \cite{wen2022rldt}.

\bibliographystyle{ieeetr}

\end{document}